\documentclass[pageno]{jpaper}

\usepackage{mathptmx} 
\usepackage{fancyhdr}
\usepackage[normalem]{ulem}
\usepackage{flushend}
\usepackage[english]{babel}
\usepackage[utf8]{inputenc}
\usepackage{algorithm}
\usepackage{float}
\usepackage[noend]{algpseudocode}
\usepackage{textcomp}
\usepackage[dvipsnames]{xcolor}
\usepackage{xspace}
\usepackage{subcaption}
\usepackage{booktabs}
\usepackage{enumitem}
\usepackage[noadjust]{cite}
\usepackage{graphicx}
\usepackage{pifont}
\usepackage{xargs} 

\usepackage{setspace}
\usepackage{mwe}
\usepackage{multirow}
\usepackage[compact]{titlesec}
\usepackage{amsmath}
\usepackage{tikz}

\usepackage[acronym,nonumberlist,nowarn]{glossaries}
    \glsdisablehyper
     \newacronym{ADI}{ADI}{Alternating Direction Implicit}
\newacronym{AGU}{AGU}{Address Generation Unit}
\newacronym[longplural={Access History Tables}]{AHT}{AHT}{Access History Table}
\newacronym{AHT-R}{AHT-R}{Access History Table - Read}
\newacronym{AHT-W}{AHT-W}{Access History Table - Write-Back}
\newacronym{AIP}{AIP}{Access Interval Predictor}
\newacronym{AL}{AL}{Added Latency for column accesses}
\newacronym{ASIC}{ASIC}{Application Specific Integrated Circuit}
\newacronym{AVX}{AVX}{Advanced Vector Extensions}
\newacronym[longplural={Basic Block Vectors}]{BBV}{BBV}{Basic Block Vector}
\newacronym{BHT}{BHT}{Branch History Table}
\newacronym{HBM}{HBM}{Hybrid Bandwidth Memory}
\newacronym{DDR4}{DDR4}{Double-Data Rate 4}
\newacronym{MIMD}{MIMD}{multiple-instruction multiple-data}
\newacronym{SIMT}{SIMT}{Single Instruction Multiple Threads}
\newacronym{SLP}{SLP}{subarray-level parallelism}
\newacronym{BLP}{BLP}{bank-level parallelism}

\newacronym{NN}{NN}{neural network}
\newacronym{BTB}{BTB}{Branch Target Buffer}
\newacronym{CAS}{CAS}{Column Address Strobe}
\newacronym{AMAT}{AMAT}{Average Memory Access Time}
\newacronym{CCD}{CCD}{Column to Column Delay}
\newacronym{AI}{AI}{Arithmetic Intensity}
\newacronym[longplural={Columnar Database Systems}]{CDBS}{CDBS}{Columnar Database System}
\newacronym[longplural={Chip Multiprocessors}]{CMP}{CMP}{Chip Multiprocessor}
\newacronym{CMT}{CMT}{Chip Multithreading}
\newacronym[longplural={Coarse-Grain Reconfigurable Arrays}]{CGRA}{CGRA}{Coarse-Grain Reconfigurable Array}
\newacronym{C-RAM}{C-RAM}{Computational-RAM}
\newacronym{CWD}{CWD}{Column Write Delay}
\newacronym{DBI}{DBI}{Dynamic Binary Instrumentation}
\newacronym[longplural={Database Management Systems}]{DBMS}{DBMS}{Database Management System}
\newacronym{DDR}{DDR}{Double Data Rate}
\newacronym{DEWP}{DEWP}{Dead Line and Early Write-Back Predictor}
\newacronym{DRAM}{DRAM}{Dynamic Random Access Memory}
\newacronym{DSBP}{DSBP}{Dead Sub-Block Predictor}
\newacronym{DSM}{DSM}{Decomposed Storage Manage}
\newacronym{FAW}{FAW}{Four row Activation Window}
\newacronym{FP}{FP}{Floating-Point}
\newacronym{EDP}{EDP}{Energy-Delay Product}
\newacronym{FCFS}{FCFS}{First-Come First-Serve}
\newacronym{FIFO}{FIFO}{Fist In, First Out}
\newacronym{FSM}{FSM}{finite state machine}
\newacronym{FPGA}{FPGA}{Field-Programmable Gate Array}
\newacronym[longplural={Functional Units}]{FU}{FU}{Functional Unit}
\newacronym{GAg}{GAg}{Global Adaptive branch prediction using one Global PHT}
\newacronym{GAs}{GAs}{Global Adaptive branch prediction using per-Set PHT}
\newacronym{GCC}{GCC}{GNU Compiler Collection}
\newacronym{GEMS}{GEMS}{General Execution-driven Multiprocessor Simulator}
\newacronym[longplural={General Purpose Processors}]{GPP}{GPP}{General Purpose Processor}
\newacronym{HMC}{HMC}{Hybrid Memory Cube}
\newacronym{HIVE}{HIVE}{HMC Instruction Vector Extensions}
\newacronym{IATAC}{IATAC}{Inter-Access Time per Access Count}
\newacronym{ILP}{ILP}{Instruction Level Parallelism}
\newacronym{IPC}{IPC}{Instructions per Cycle}
\newacronym{ISA}{ISA}{Instruction Set Architecture}
\newacronym{IRAM}{IRAM}{Intelligent RAM}
\newacronym{KIPS}{KIPS}{Kilo Instructions per Second}
\newacronym{LDS}{LDS}{Linked Data Structure}
\newacronym{LFSR}{LFSR}{Linear Feedback Shift Register}
\newacronym{LFMR}{LFMR}{Last-to-First Miss-Ratio}
\newacronym{MLP}{MLP}{Memory-Level Parallelism}

\newacronym{LLC}{LLC}{last-level cache}
\newacronym{LRU}{LRU}{Least Recently Used}
\newacronym{LSU}{LSU}{Load Store Unit}
\newacronym{LTP}{LTP}{Last-Touch Predictor}
\newacronym{LvP}{LvP}{Live-time Predictor}
\newacronym{LWP}{LWP}{Last Write Predictor}
\newacronym{MARSS}{MARSS}{Micro Architectural and System Simulator}
\newacronym{McPAT}{McPAT}{Multi-core Power, Area, and Timing}
\newacronym{MIPS}{MIPS}{Microprocessor without Interlocked Pipeline Stages}
\newacronym{MOB}{MOB}{Memory Order Buffer}
\newacronym{MMU}{MMU}{memory management unit}
\newacronym{PuD}{PUD}{processing-using-DRAM}
\newacronym{MPKI}{MPKI}{Misses per Kilo-Instruction}
\newacronym{MSHR}{MSHR}{Miss-Status Handling Registers}
\newacronym{NAS}{NAS}{Numerical Aerodynamic Simulation}
\newacronym{NDP}{NDP}{Near-Data Processing}
\newacronym{NMP}{NMP}{Near Memory Processor}
\newacronym{NoC}{NoC}{Network-on-Chip}
\newacronym{NPB}{NPB}{NAS Parallel Benchmark}
\newacronym{NUCA}{NUCA}{Non-Uniform Cache Architecture}
\newacronym{NUMA}{NUMA}{Non-Uniform Memory Access}
\newacronym{OoO}{OoO}{Out-of-Order}
\newacronym{OpenMP}{OpenMP}{Open Multi-Processing}
\newacronym{OS}{OS}{operating system}
\newacronym{PAg}{PAg}{Per-address Adaptive branch prediction using one Global PHT}
\newacronym{PAs}{PAs}{Per-address Adaptive branch prediction using per-Set PHT}
\newacronym{PC}{PC}{Program Counter}
\newacronym[longplural={Pointer-Chasing Engines}]{PCE}{PCE}{Pointer-Chasing Engine}
\newacronym{PCM}{PCM}{Performance Counter Monitor}
\newacronym{PIM}{PIM}{processing-in-memory}
\newacronym[longplural={Pattern History Tables}]{PHT}{PHT}{Pattern History Table}
\newacronym{RAPL}{RAPL}{Running Average Power Limit}
\newacronym{RAS}{RAS}{Row Address Strobe}
\newacronym{RAT}{RAT}{Registers Alias Table}
\newacronym{RC}{RC}{Row Cycle}
\newacronym{RCD}{RCD}{RAS to CAS Delay}
\newacronym[longplural={Row-based Database Systems}]{RDBS}{RDBS}{Row-based Database System}
\newacronym{ROB}{ROB}{Reorder Buffer}
\newacronym{RRD}{RRD}{Row to Row activation Delay}
\newacronym{DNN}{DNN}{Deep Neural Network}
\newacronym{ANN}{ANN}{Artificial Neural Network}
\newacronym{PnM}{PNM}{processing-near-memory}
\newacronym{PuM}{PUM}{processing-using-memory}
\newacronym{FFD}{FFD}{first-fit-decreasing}
\newacronym{HFF}{HFF}{helper flip-flop}
\newacronym{OBPS}{OBPS}{one-bit per-subarray}
\newacronym{GEMV}{GEMV}{general matrix-vector product}

\newacronym{RBR}{RBR}{redundant binary representation}
\newacronym{RP}{RP}{Row Precharge}
\newacronym{RTL}{RTL}{Register Transfer Level}
\newacronym{RTP}{RTP}{Read To Precharge}
\newacronym{RVU}{RVU}{Reconfigurable Vector Unit}
\newacronym{SDP}{SDP}{Skewed Dead-Block Predictor}
\newacronym{SESC}{SESC}{Superescalar Simulator}
\newacronym{SSE}{SSE}{Streaming SIMD Extensions}
\newacronym{SFP}{SFP}{Spatial Footprint Predictor}
\newacronym{SiNUCA}{SiNUCA}{Simulator of Non-Uniform Cache Architectures}
\newacronym{SMT}{SMT}{Simultaneous Multi-Threading}
\newacronym{SIMD}{SIMD}{single-instruction multiple-data}
\newacronym{LUT}{LUT}{lookup table}

\newacronym{SPEC}{SPEC}{Standard Performance Evaluation Corporation}
\newacronym{SoC}{SoC}{System-on-Chip}
\newacronym{SPP}{SPP}{Spatial Pattern Predictor}
\newacronym{SRAM}{SRAM}{Static Random Access Memory}
\newacronym{SSOR}{SSOR}{Symmetric Successive Over-Relaxation}
\newacronym{SSV}{SSV}{Search Set Vector}
\newacronym{TLB}{TLB}{Translation Look-aside Buffer}
\newacronym{TLP}{TLP}{Thread Level Parallelism}
\newacronym[longplural={Through-Silicon Vias}]{TSV}{TSV}{Through-Silicon Via}
\newacronym{VWQ}{VWQ}{Virtual Write Queue}
\newacronym{WTR}{WTR}{Write Recovery time}
\newacronym{WR}{WR}{Write To Read delay time}
\newacronym{TRA}{TRA}{triple row activation}

\newcommand{\circled}[1]{\tikz[baseline=(char.base)]{\node[shape=circle,draw,inner sep=0pt,fill=black, text=white] (char) {#1};}}

\makeatletter

\renewcommand\newblock{\hskip .11em\@plus.33em\@minus.07em}
\makeatother

\newcommand{\li}{(\textit{i})}
\newcommand{\lii}{(\textit{ii})}
\newcommand{\liii}{(\textit{iii})}

\newcommand{\prop}{PUMA\xspace}

\newcommand{\paratitle}[1]{\vspace{4pt}\noindent\textbf{#1.}}

\newcommand{\circledii}[1]{\tikz[baseline=(char.base)]{\node[shape=circle,draw,inner sep=0pt,fill=gray, text=white] (char) {\itshape#1};}}

\definecolor{airforceblue}{rgb}{0.36, 0.54, 0.66}
\definecolor{dodgerblue}{rgb}{0.12, 0.56, 1.0}
\definecolor{brandeisblue}{rgb}{0.0, 0.44, 1.0}
\definecolor{brickred}{rgb}{0.8, 0.25, 0.33}
\definecolor{eggplant}{rgb}{0.38, 0.25, 0.32}
\definecolor{byzantium}{rgb}{0.44, 0.16, 0.39}
\definecolor{ddgreen}{rgb}{0.00, 0.50, 0.00}

\definecolor{mygreen}{rgb}{0,0.6,0}
\definecolor{mygray}{rgb}{0.5,0.5,0.5}
\definecolor{mymauve}{rgb}{0.58,0,0.82}

\definecolor{bluehl}{rgb}{0.8,0.874,1}
\definecolor{pinkhl}{rgb}{0.992156863,0.847058824,1}
\definecolor{macaroniandcheese}{rgb}{1.0, 0.74, 0.53}
\definecolor{mossgreen}{rgb}{0.68, 0.87, 0.68}
\definecolor{greenhl}{rgb}{0.835,0.996,0.939}
\definecolor{yellowhl}{rgb}{0.996,0.957,0.8}
\definecolor{palecerulean}{rgb}{0.61, 0.77, 0.89}
\definecolor{gray(x11gray)}{rgb}{0.75, 0.75, 0.75}

\newcommand\ignore[1]{ }
\newcommand{\revdel}[1]{}

\newif\ifcut
\cutfalse

\ifcut
   \newcommand{\gfcut}[1]{} 
\else
    \newcommand{\gfcut}[1]{\textcolor{red}{\sout{#1}}}

\fi 

\newif\ifsubmission
\submissiontrue

\ifsubmission

    \newcommand{\jgl}[1]{}

    \newcommand{\gfb}[1]{}
    \newcommand{\mayank}[1]{}
    
    \newcommand{\agy}[1]{#1}
    \newcommand{\agycomment}[1]{}
\else
    \newcommand{\jgl}[1]{\textcolor{brickred}{\textit{JGL: #1}}}
    \newcommand{\gfb}[1]{\textcolor{blue}{\textit{GF: #1}}}

    \newcommand{\mayank}[1]{\textcolor{green}{\textit{Mayank: #1}}}

    \newcommand{\agy}[1]{\textcolor{orange}{#1}}
    \newcommand{\agycomment}[1]{\agy{\textbf{[@gy:} #1\textbf{]}}}

\fi

\newcommand\pimdef{\cite{ghose.ibmjrd19, mutlu2020modern,deoliveira2021IEEE,pim-book,mutlu2019processing,mutlu2019enabling,mutlu2015research,mutlu2013memory,loh2013processing,Near-Data,stone1970logic,Miss_Mem_Wall_1996}\xspace}

\newcommand\pnm{\cite{farmahini2015nda,babarinsa2015jafar,devaux2019true,ghiasi2022genstore,gomez2021benchmarkingcut,gomezluna2021benchmarking,gomez2022benchmarking,syncron,singh2020nero,skhynixpim,ke2021near,giannoula2022sparsep,shin2018mcdram,cho2020mcdram,denzler2021casper,asghari2016chameleon,IRAM_Micro_1997,C_RAM_1999,CASES_MVX,Xi_2015,sun2021abc,matam2019graphssd,gokhale1995processing,hall1999mapping,MEMSYS_MVX,lockerman2020livia,ahn2015scalable,nai2017graphpim,boroumand2018google,lazypim, top-pim, gao2016hrl, kim2018grim, drumond2017mondrian, RVU, NIM, PEI, gao2017tetris,Kim2016,gu2016leveraging, boroumand2019conda, hsieh2016transparent, cali2020genasm, NDC_ISPASS_2014,pattnaik2016scheduling,akin2015data,hsieh2016accelerating,lee2015bssync,boroumand2021mitigating,boroumand2021google,boroumand2022polynesia,boroumand2021polynesia,amiraliphd,besta2021sisa,fernandez2020natsa,singh2019napel,kwon202125,lee2021hardware,niu2022184qps,Sparse_MM_LiM,azarkhish2016logic,azarkhish2018neurostream,guo20143d,de2018design,akin2014hamlet,huang2020heterogeneous,dai2018graphh,liu2018processing,tsai:micro:2018:ams,gu2020ipim,DRAMA_CAL_2014,Asghari-Moghaddam_2016,huang2019active,kersey2017lightweight,li2019pims,kim2017grim,boroumand2017lazypim,zhuo2019graphq,zhang2018graphp,lim2017triple,smc_sim,HIVE,jang2019charon,IBM_ActiveCube,hadidi2017cairo,santos2018processing}\xspace}

\newcommand\pum{\cite{Chi2016, Shafiee2016, seshadri2017ambit, seshadri2019dram, li2017drisa, seshadri2013rowclone, seshadri2016processing, deng2018dracc, xin2020elp2im, song2018graphr, song2017pipelayer,gao2019computedram, eckert2018neural, aga2017compute,dualitycache,besta2021sisa,seshadri2016buddy,seshadri.bookchapter17,seshadri2018rowclone,seshadri2015fast,li2016pinatubo,ferreira2021pluto,ferreira2022pluto,imani2019floatpim,he2020sparse,flashcosmos,truong2022adapting,truong2021racer,olgun2021quactrng,kim2019d,kim2018dram,bostanci2022dr,olgun2022pidram,ali2019memory,angizi2019graphide,li2018scope,subramaniyan2017parallel,zha2020hyper,fujiki2018memory,orosa2021codic,sharad2013ultra,rezaei2020nom}\xspace}

\newcommand\drampum{\cite{angizi2019graphide,besta2021sisa,bostanci2022dr,deng2018dracc,ferreira2021pluto,ferreira2022pluto,gao2019computedram,li2017drisa,li2018scope,olgun2021quactrng,olgun2022pidram,seshadri.bookchapter17, seshadri2013rowclone,seshadri2015fast,seshadri2016buddy, seshadri2016processing, seshadri2017ambit,seshadri2018rowclone, seshadri2019dram, xin2020elp2im}\xspace}

\titlespacing\section{0pt}{5pt plus 2pt minus 2pt}{0pt plus 2pt minus 2pt}
\titlespacing\subsection{0pt}{5pt plus 2pt minus 2pt}{0pt plus 2pt minus 2pt}
\titlespacing\subsubsection{0pt}{5pt plus 2pt minus 2pt}{0pt plus 2pt minus 2pt}

\makeatletter
\g@addto@macro{\normalsize}{%
  \setlength{\abovedisplayskip}{2pt plus 0.5pt minus 1pt}
  \setlength{\belowdisplayskip}{2pt plus 0.5pt minus 1pt}
  \setlength{\abovedisplayshortskip}{0pt}
  \setlength{\belowdisplayshortskip}{0pt}
  \setlength{\intextsep}{2pt plus 1pt minus 1pt}
  \setlength{\textfloatsep}{2pt plus 1pt minus 1pt}
  \setlength{\skip\footins}{5pt plus 1pt minus 1pt}}
\makeatother

\begin{document}

\bstctlcite{IEEEexample:BSTcontrol}

\title{\prop: Efficient and Low-Cost Memory Allocation and Alignment Support for Processing-Using-Memory Architectures}

\author{
        \vspace{-15pt}\\ 
        \scalebox{1.0}{Geraldo F. Oliveira}~\qquad%
        \scalebox{1.0}{Emanuele G. Esposito}~\qquad%
        \scalebox{1.0}{Juan Gómez-Luna}~\qquad%
        \scalebox{1.0}{Onur Mutlu}%
    \\%
    \vspace{-10pt}\\%
         \it\normalsize  ETH Z{\"u}rich  
    \vspace{-20pt}%
}

\date{}
\maketitle

\section{Motivation \& Problem}

\Gls{PIM}~\pimdef is a promising paradigm that aims to alleviate the ever-growing cost of moving data back and forth between computing (e.g., CPU, GPU, accelerators) and memory (e.g., caches, main memory, storage) elements. 
In \gls{PIM} architectures, computation is done by adding logic units \emph{near} memory arrays, i.e., \gls{PnM}~\pnm, or by \emph{using} the analog properties of the memory arrays themselves, i.e., \gls{PuM}~\pum). 
Several prior works~\drampum have demonstrated the feasibility of \gls{PuD} architectures, which use DRAM cells to implement a variety of \gls{PuM} operations, including data copy and initilization~\cite{seshadri2013rowclone, seshadri2018rowclone}, bitwise Boolean~\cite{seshadri2017ambit, gao2019computedram, xin2020elp2im, besta2021sisa, li2017drisa}, and arithmetic operations~\cite{deng2018dracc, gao2019computedram,li2017drisa,angizi2019graphide, hajinazarsimdram,li2018scope}. 

\gls{PuD} architectures impose a restrictive data layout and alignment for their operands, where source and destination operands 
\li~\emph{must}  reside in the same DRAM subarray (i.e., a group of DRAM rows sharing the same row buffer and row decoder) and 
\lii~are aligned to the boundaries of a DRAM row. 
However, standard memory allocation routines (i.e., \texttt{malloc}, \texttt{posix\_memalign}, and \texttt{huge pages}-based memory allocation) fail to meet the data layout and alignment requirements for \gls{PuD} architectures to operate successfully for two main reasons. 
First, while \texttt{malloc} and \texttt{posix\_memalign} can provide the user application virtually aligned contiguous memory pages, they do \emph{not} guarantee that the allocated virtual pages are also contiguous in physical memory and aligned within a DRAM row. 
Second, employing \texttt{huge pages}-based memory allocation can guarantee that virtual pages are contiguous in physical memory. However, due to its \emph{coarse-grained} page allocation sizes (i.e., Linux-based systems can provide huge pages of 2~MB or 1~GB), a \emph{single} huge page allocation can cover \emph{all} the rows in a DRAM subarray in a single DRAM chip.\footnote{A typical DRAM subarray has 1024 DRAM rows, each with 1024 DRAM columns. Thus, a single DRAM subarray can store 1~MB of data.~\cite{}} Therefore, when the \gls{PuD} instruction requires multiple operands (and thus multiple huge page allocations), it is likely that such operands will resign in different DRAM subarrays, thus imposing extra latency due to inter-subarray data movement~\cite{chang2016low}.

We investigate the potential of using \texttt{malloc}, \texttt{posix\_memalign}, and \texttt{huge pages}-based memory allocation for a \gls{PuD} substrate that can execute AND/OR/NOT Boolean operations (i.e., Ambit~\cite{seshadri2017ambit}). We consider that AND/OR/NOT Boolean operations can be executed in the \gls{PuD} substrate \emph{only} when the data alignment and allocation requirements are met (i.e., source and destination operands are contiguous in physical memory and DRAM row-aligned). We observe that 
(i)~independently of the allocation size for input operands, using \texttt{malloc} and \texttt{posix\_memalign} memory allocators results in 0\% of the operations being executed in the \gls{PuD} substrate due to data misalignment; and
(ii)~for large-enough allocation sizes (e.g., 32~Kb), \emph{only} up 60\% of the \gls{PuD} operations that use \texttt{huge pages}-based memory allocation can successfully be executed in DRAM. We conclude that traditional memory allocators cannot take full advantage of such \gls{PuD} techniques since they cannot satisfy the specific memory allocation requirements of \gls{PuD} substrates. Therefore, our \textbf{goal} of this work is to provide a flexible memory allocation mechanism that allows programmers to have control over physical memory allocation and enables \gls{PuD} execution from the \gls{OS} viewpoint.

\section{\prop: Key Idea \& Overview}

\begin{figure*}
    \centering
    \includegraphics[width=0.95\textwidth]{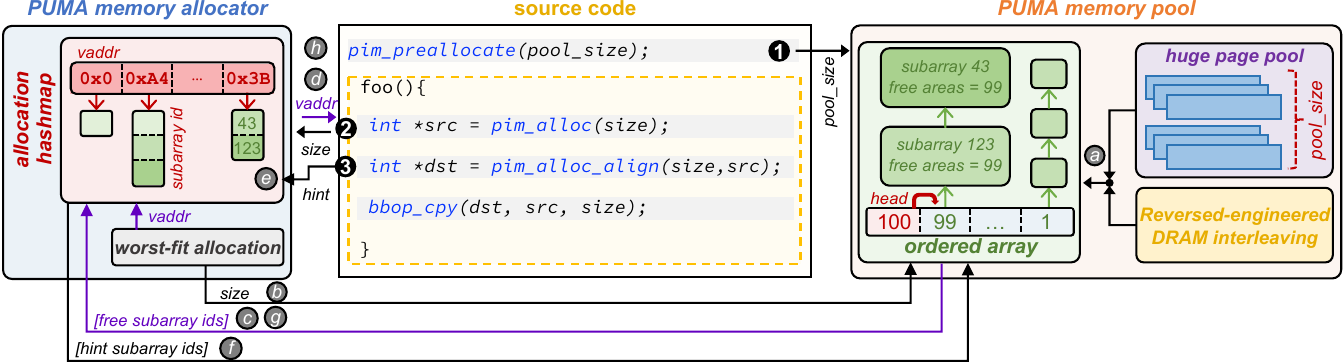}
    \caption{Overview of the \prop framework.}
    \label{fig:framework_overview}
\end{figure*}

To allow the memory allocation API to influence the \gls{OS} memory allocator and ensure that memory objects are placed within specific DRAM subarrays, we propose a new \emph{lazy data allocation routine} (in the kernel) for \gls{PuD} memory objects, called \prop.  
The \emph{key idea} of \prop is to use the internal DRAM mapping information, together with huge pages, and then split huge pages into \emph{finer-grained} allocation units that are 
\li~aligned to the page address and size and 
\lii~virtually contiguous. The \prop routine has three main components (as Figure~\ref{fig:framework_overview} illustrates): 
\li~information regarding the DRAM organization (e.g., row, column, and mat sizes), 
\lii~the DRAM interleaving scheme, which the memory controller provides via an open firmware device tree~\cite{devicethree};\footnote{The DRAM interleaving scheme can be obtained by reverse engineering the bit locations of memory addresses~\cite{kim2020revisiting, orosa2021deeper,yauglikcci2022understanding}.} and 
\liii~a huge pages pool for \gls{PuD} memory objects (configured during boot time), which guarantees that virtual addresses assigned to a \gls{PuD} memory objects are contiguous in the physical address space.  
The allocation routine uses the DRAM address mapping knowledge to split the huge pages into different memory regions. Then, it uses the DRAM interleaving scheme to index each memory region based on their subarray ID (obtained by ORing subarray, bank, channel, and rank mask bits in the DRAM interleaving scheme). \prop uses an \emph{ordered array} data structure similar to the one used in the Linux Kernel buddy allocator algorithm~\cite{knowlton1966programmer}, where each entry represents the number of memory regions in a single subarray. When an application calls the  \gls{PuD} memory allocation API, the allocation routine selects the appropriate memory region that satisfies the memory allocation.  \prop operates by exposing three new memory allocation APIs to the user: 
\li~\texttt{pim\_preallocate}, for pre-allocation;
\lii~\texttt{pim\_alloc}, for the first data allocation; and
\liii~\texttt{pim\_alloc\_align}, for subsequent aligned allocations. 

\paratitle{Pre-Allocation} The first step in \prop is to indicate the number of huge pages that are available for \gls{PuD} allocations using the \texttt{pim\_preallocate} API (\circled{1} in Figure~\ref{fig:framework_overview}). We left the user the duty to provide the number of huge pages used for \gls{PuD} operations (\circledii{a} in Figure~\ref{fig:framework_overview}) because huge pages are scarce in the system. 

\paratitle{First Allocation} \prop uses the \emph{worst-fit allocation scheme}~\cite{johnson1973near} to manage the allocation of memory regions in the huge page pool. The main idea behind this placement strategy is to optimize the remaining space post-allocations, thereby increasing the chances of accommodating another process in the remaining memory space. Based on that, for the first \gls{PuD} memory allocation (using the \texttt{pim\_alloc} API; \circled{2}), \prop simply scans the \emph{ordered array} to select the subarray with the \emph{largest} amount of memory regions available (\circledii{b}). If the requested memory allocation requires more than one memory region, \prop interactively scans the  \emph{ordered array}, searching for the next largest memory region until the memory allocation is fully satisfied. Once enough space is allocated (\circledii{c}), \prop creates a new allocation object and inserts it in an \emph{allocation hashmap}, which is indexed (\circledii{d}) by the allocation's virtual address. \prop needs to keep track of allocations since it might need to find a memory region from the \emph{same} subarray when performing the future aligned allocations (i.e., for the second operand for a Boolean operation). 

\paratitle{Aligned Allocation} After allocating memory regions for the first operand in a \gls{PuD} operation, the user can use this memory region as a regular memory object. However, when allocating the remaining operands for a \gls{PuD} operations (e.g., the second input operand and destination operand in a vector-based Boolean AND operation), \prop needs to guarantee data alignment for all memory objects within the same DRAM subarray. To this end, we implement a new memory allocation API called \texttt{pim\_alloc\_align}, which takes a \texttt{hint} pointer as input (\circled{3}). Such a pointer indicates a previously allocated memory region to which the current memory allocation must be aligned. The \texttt{pim\_alloc\_align} allocation API works in five main steps. 
First, \prop searches the \emph{allocation hashmap} for a match with the address in the \texttt{hint} pointer (\circledii{e}). If a match is not found, the allocation fails.
Second, if a match is found, \prop iterates through the \texttt{hint}-allocation's memory regions (\circledii{f}). 
Third, for each memory region, \prop identifies its source subarray address and tries to allocate another memory region at the same subarray for the new allocation (\circledii{g}).
Fourth, if the subarray of a given memory region has no free region, \prop allocates a new memory region from another subarray following the worst-fit allocation scheme (\circledii{h}). Since we use a worst-fit allocation scheme, we have a good chance of having a single subarray holding memory regions for multiple allocations.
Fifth, since memory regions might come from different huge pages, we must perform \texttt{re-mmap} to map such memory regions into contiguous virtual addresses. 

\section{Key Results \& Contributions}

\paratitle{Evaluation Methodology} We implement \prop as a kernel module using QEMU~\cite{qemu}, an open-source emulator and virtualize that can perform hardware virtualization. We emulate a RISC-V machine running Fedora 33 with v5.9.0 Linux Kernel. In our experiments, we evaluate a system with 8~GB DRAM. We emulate the implementation of a \gls{PuD} system capable of executing row copy operations (as in RowClone~\cite{seshadri2013rowclone}) and Boolean AND/OR/NOT operations (as in Ambit~\cite{seshadri2017ambit}).  In our experiments, such an operation is performed in the host CPU if a given operation cannot be executed in our \gls{PuD} substrate (due to data misalignment).

\paratitle{Baselines \& Workloads} We compare the performance of \prop to that of using traditional CPU \texttt{memcpy} allocation.\footnote{\texttt{posix\_mem\_align} shows the same performance as \texttt{memcpy}.} We use three micro-benchmarks in our analysis:
\li~initialize an array with zeros (\texttt{*-zero}),
\lii~copy data from one array to another (\texttt{*-copy});
\liii~perform vector bitwise AND operations \texttt{C[i] = A[i] AND B[i]} using Ambit (\texttt{*-aand}). For each micro-benchmark, we vary the allocation sizes from 2000~bits to 6~Mb. 

\begin{figure}[ht]
    \centering
    \includegraphics[width=\linewidth]{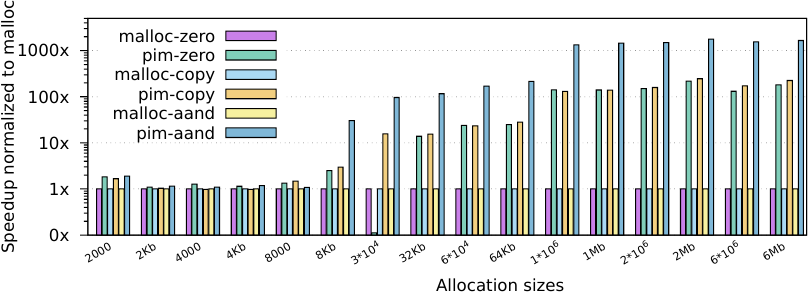}
    \caption{\prop's performance for three micro-benchmarks and varying data allocation sizes. Values are normalized to the baseline \texttt{malloc} allocator.}
    \label{fig:results}
\end{figure}

\paratitle{Evaluation Results} Figure~\ref{fig:results} shows \prop's performance for each micro-benchmark for different allocation sizes (x-axis) compared to the baseline \texttt{malloc} allocator (y-axis). We make two observations for the figure. 
First, \prop \emph{significantly} outperforms the baseline memory allocators for all micro-benchmarks and allocation sizes. This is because \prop increases the likelihood of an operation to be executed in DRAM (due to proper data alignment and allocation), thus increasing overall performance. 
Second, \prop's performance improvements increase as the data allocation sizes increase. This is because the larger the allocation, the more data would need to be moved from DRAM to the CPU in case a \gls{PuD} operation fails to be executed. Thus severely penalizing overall performance. We conclude that \prop is a practical and efficient memory allocator for \gls{PuD} substrates.

We make the following key contributions:
\begin{itemize}[noitemsep,topsep=0pt,parsep=0pt,partopsep=0pt,labelindent=0pt,itemindent=0pt,leftmargin=*] 
    \item To our knowledge, this is the first work to propose a practical memory allocation mechanism for \gls{PuD} substrates. 

    \item We propose \prop, a data allocation routine for \gls{PuD} architectures that use the internal DRAM mapping information and huge pages to provide aligned data allocation for \gls{PuD} instructions.

    \item \prop does \emph{not} require hardware modifications and operates transparently from the user as a Linux kernel module. 

    \item We evaluate \prop using three micro-benchmarks, and we observe that \prop \emph{significantly} increases performance compared to \texttt{malloc}-based memory allocators.
\end{itemize}

\footnotesize
\bibliographystyle{IEEEtran}
\bibliography{refs}

\begin{thebibliography}{100}
\providecommand{\url}[1]{#1}
\csname url@samestyle\endcsname
\providecommand{\newblock}{\relax}
\providecommand{\bibinfo}[2]{#2}
\providecommand{\BIBentrySTDinterwordspacing}{\spaceskip=0pt\relax}
\providecommand{\BIBentryALTinterwordstretchfactor}{4}
\providecommand{\BIBentryALTinterwordspacing}{\spaceskip=\fontdimen2\font plus
\BIBentryALTinterwordstretchfactor\fontdimen3\font minus \fontdimen4\font\relax}
\providecommand{\BIBforeignlanguage}[2]{{%
\expandafter\ifx\csname l@#1\endcsname\relax
\typeout{** WARNING: IEEEtran.bst: No hyphenation pattern has been}%
\typeout{** loaded for the language `#1'. Using the pattern for}%
\typeout{** the default language instead.}%
\else
\language=\csname l@#1\endcsname
\fi
#2}}
\providecommand{\BIBdecl}{\relax}
\BIBdecl

\bibitem{ghose.ibmjrd19}
S.~Ghose, A.~Boroumand, J.~S. Kim, J.~G{\'o}mez-Luna, and O.~Mutlu, ``{Processing-in-Memory: A Workload-Driven Perspective},'' \emph{IBM JRD}, 2019.

\bibitem{mutlu2020modern}
O.~Mutlu, S.~Ghose, J.~G{\'o}mez-Luna, and R.~Ausavarungnirun, ``{A Modern Primer on Processing in Memory},'' in \emph{Emerging Computing: From Devices to Systems --- Looking Beyond Moore and Von Neumann}.\hskip 1em plus 0.5em minus 0.4em\relax Springer, 2021.

\bibitem{deoliveira2021IEEE}
G.~F. Oliveira, J.~Gómez-Luna, L.~Orosa, S.~Ghose, N.~Vijaykumar, I.~Fernandez, M.~Sadrosadati, and O.~Mutlu, ``{DAMOV: A New Methodology and Benchmark Suite for Evaluating Data Movement Bottlenecks},'' \emph{IEEE Access}, 2021.

\bibitem{pim-book}
S.~Ghose, K.~Hsieh, A.~Boroumand, R.~Ausavarungnirun, and O.~Mutlu, ``{The Processing-in-Memory Paradigm: Mechanisms to Enable Adoption},'' in \emph{Beyond-CMOS Technologies for Next Generation Computer Design}, 2019.

\bibitem{mutlu2019processing}
O.~Mutlu, S.~Ghose, J.~G{\'o}mez-Luna, and R.~Ausavarungnirun, ``{Processing Data Where It Makes Sense: Enabling In-Memory Computation},'' \emph{MicPro}, 2019.

\bibitem{mutlu2019enabling}
O.~Mutlu, S.~Ghose, J.~G{\'o}mez-Luna, and R.~Ausavarungnirun, ``{Enabling Practical Processing in and Near Memory for Data-Intensive Computing},'' in \emph{DAC}, 2019.

\bibitem{mutlu2015research}
O.~Mutlu and L.~Subramanian, ``{Research Problems and Opportunities in Memory Systems},'' \emph{SUPERFRI}, 2014.

\bibitem{mutlu2013memory}
O.~Mutlu, ``{Memory Scaling: A Systems Architecture Perspective},'' in \emph{IMW}, 2013.

\bibitem{loh2013processing}
G.~H. Loh, N.~Jayasena, M.~Oskin, M.~Nutter, D.~Roberts, M.~Meswani, D.~P. Zhang, and M.~Ignatowski, ``{A Processing in Memory Taxonomy and a Case for Studying Fixed-Function PIM},'' in \emph{WoNDP}, 2013.

\bibitem{Near-Data}
R.~Balasubramonian, J.~Chang, T.~Manning \emph{et~al.}, ``{Near-Data Processing: Insights from a MICRO-46 Workshop},'' \emph{IEEE Micro}, 2014.

\bibitem{stone1970logic}
H.~S. Stone, ``{A Logic-in-Memory Computer},'' \emph{IEEE Trans. Comput.}, 1970.

\bibitem{Miss_Mem_Wall_1996}
A.~Saulsbury, F.~Pong, and A.~Nowatzyk, ``{Missing the Memory Wall: The Case for Processor/Memory Integration},'' in \emph{ISCA}, 1996.

\bibitem{farmahini2015nda}
A.~Farmahini-Farahani, J.~H. Ahn, K.~Morrow, and N.~S. Kim, ``{NDA: Near-DRAM Acceleration Architecture Leveraging Commodity DRAM Devices and Standard Memory Modules},'' in \emph{HPCA}, 2015.

\bibitem{babarinsa2015jafar}
O.~O. Babarinsa and S.~Idreos, ``{JAFAR: Near-Data Processing for Databases},'' in \emph{SIGMOD}, 2015.

\bibitem{devaux2019true}
F.~Devaux, ``{The True Processing in Memory Accelerator},'' in \emph{Hot Chips}, 2019.

\bibitem{ghiasi2022genstore}
N.~M. Ghiasi, J.~Park, H.~Mustafa, J.~Kim, A.~Olgun, A.~Gollwitzer, D.~S. Cali, C.~Firtina, H.~Mao, N.~A. Alserr \emph{et~al.}, ``{GenStore: A High-Performance and Energy-Efficient In-Storage Computing System for Genome Sequence Analysis},'' in \emph{ASPLOS}, 2022.

\bibitem{gomez2021benchmarkingcut}
J.~G{\'o}mez-Luna, I.~El~Hajj, I.~Fernandez, C.~Giannoula, G.~F. Oliveira, and O.~Mutlu, ``{Benchmarking Memory-Centric Computing Systems: Analysis of Real Processing-in-Memory Hardware},'' in \emph{CUT}, 2021.

\bibitem{gomezluna2021benchmarking}
J.~G{\'o}mez-Luna, I.~E. Hajj, I.~Fernández, C.~Giannoula, G.~F. Oliveira, and O.~Mutlu, ``{Benchmarking a New Paradigm: An Experimental Analysis of a Real Processing-in-Memory Architecture},'' arXiv:2105.03814 [cs.AR], 2021.

\bibitem{gomez2022benchmarking}
J.~G{\'o}mez-Luna, I.~El~Hajj, I.~Fernandez, C.~Giannoula, G.~F. Oliveira, and O.~Mutlu, ``{Benchmarking a New Paradigm: Experimental Analysis and Characterization of a Real Processing-in-Memory System},'' \emph{IEEE Access}, 2022.

\bibitem{syncron}
C.~Giannoula, N.~Vijaykumar, N.~Papadopoulou, V.~Karakostas, I.~Fernandez, J.~Gómez-Luna, L.~Orosa, N.~Koziris, G.~Goumas, and O.~Mutlu, ``{SynCron: Efficient Synchronization Support for Near-Data-Processing Architectures},'' in \emph{HPCA}, 2021.

\bibitem{singh2020nero}
G.~Singh, D.~Diamantopoulos, C.~Hagleitner, J.~Gomez-Luna, S.~Stuijk, O.~Mutlu, and H.~Corporaal, ``{NERO: A Near High-Bandwidth Memory Stencil Accelerator for Weather Prediction Modeling},'' in \emph{FPL}, 2020.

\bibitem{skhynixpim}
S.~Lee, K.~Kim, S.~Oh, J.~Park, G.~Hong, D.~Ka, K.~Hwang, J.~Park, K.~Kang, J.~Kim, J.~Jeon, N.~Kim, Y.~Kwon, K.~Vladimir, W.~Shin, J.~Won, M.~Lee, H.~Joo \emph{et~al.}, ``{A 1ynm 1.25V 8Gb, 16Gb/s/pin GDDR6-based Accelerator-in-Memory Supporting 1TFLOPS MAC Operation and Various Activation Functions for Deep-Learning Applications},'' in \emph{ISSCC}, 2022.

\bibitem{ke2021near}
L.~Ke, X.~Zhang, J.~So, J.-G. Lee, S.-H. Kang, S.~Lee, S.~Han, Y.~Cho, J.~H. Kim, Y.~Kwon \emph{et~al.}, ``{Near-Memory Processing in Action: Accelerating Personalized Recommendation with AxDIMM},'' \emph{IEEE Micro}, 2021.

\bibitem{giannoula2022sparsep}
C.~Giannoula, I.~Fernandez, J.~G. Luna, N.~Koziris, G.~Goumas, and O.~Mutlu, ``{SparseP: Towards Efficient Sparse Matrix Vector Multiplication on Real Processing-in-Memory Architectures},'' in \emph{SIGMETRICS}, 2022.

\bibitem{shin2018mcdram}
H.~Shin, D.~Kim, E.~Park, S.~Park, Y.~Park, and S.~Yoo, ``{McDRAM: Low Latency and Energy-Efficient Matrix Computations in DRAM},'' \emph{IEEE TCADICS}, 2018.

\bibitem{cho2020mcdram}
S.~Cho, H.~Choi, E.~Park, H.~Shin, and S.~Yoo, ``{McDRAM v2: In-Dynamic Random Access Memory Systolic Array Accelerator to Address the Large Model Problem in Deep Neural Networks on the Edge},'' \emph{IEEE Access}, 2020.

\bibitem{denzler2021casper}
A.~Denzler, R.~Bera, N.~Hajinazar, G.~Singh, G.~F. Oliveira, J.~G{\'o}mez-Luna, and O.~Mutlu, ``{Casper: Accelerating Stencil Computation using Near-Cache Processing},'' arXiv:2112.14216 [cs.AR], 2021.

\bibitem{asghari2016chameleon}
H.~Asghari-Moghaddam, Y.~H. Son, J.~H. Ahn, and N.~S. Kim, ``{Chameleon: Versatile and Practical Near-DRAM Acceleration Architecture for Large Memory Systems},'' in \emph{MICRO}, 2016.

\bibitem{IRAM_Micro_1997}
D.~Patterson, T.~Anderson, N.~Cardwell \emph{et~al.}, ``{A Case for Intelligent RAM},'' \emph{IEEE Micro}, 1997.

\bibitem{C_RAM_1999}
D.~G. Elliott, M.~Stumm, W.~M. Snelgrove \emph{et~al.}, ``{Computational RAM: Implementing Processors in Memory},'' \emph{Design and Test of Computers}, vol.~16, no.~1, pp. 32--41, Jan. 1999.

\bibitem{CASES_MVX}
M.~A.~Z. Alves, P.~C. Santos, F.~B. Moreira, and opthers, ``Saving memory movements through vector processing in the dram,'' in \emph{Int. Conf. on Compilers, Architecture and Synthesis for Embedded Systems}, 2015.

\bibitem{Xi_2015}
S.~L. Xi, O.~Babarinsa, M.~Athanassoulis, and S.~Idreos, ``Beyond the wall: Near-data processing for databases,'' in \emph{Int. Workshop on Data Management on New Hardware}, 2015.

\bibitem{sun2021abc}
W.~Sun, Z.~Li, S.~Yin, S.~Wei, and L.~Liu, ``{ABC-DIMM: Alleviating the Bottleneck of Communication in DIMM-Based Near-Memory Processing with Inter-DIMM Broadcast},'' in \emph{ISCA}, 2021.

\bibitem{matam2019graphssd}
K.~K. Matam, G.~Koo, H.~Zha, H.-W. Tseng, and M.~Annavaram, ``{GraphSSD: Graph Semantics Aware SSD},'' in \emph{ISCA}, 2019.

\bibitem{gokhale1995processing}
M.~Gokhale, B.~Holmes, and K.~Iobst, ``{Processing in Memory: The Terasys Massively Parallel PIM Array},'' \emph{Computer}, 1995.

\bibitem{hall1999mapping}
M.~Hall, P.~Kogge, J.~Koller, P.~Diniz, J.~Chame, J.~Draper, J.~LaCoss, J.~Granacki, J.~Brockman, A.~Srivastava \emph{et~al.}, ``{Mapping Irregular Applications to DIVA, a PIM-Based Data-Intensive Architecture},'' in \emph{SC}, 1999.

\bibitem{MEMSYS_MVX}
M.~A.~Z. Alves, P.~C. Santos, M.~Diener, and L.~Carro, ``{Opportunities and Challenges of Performing Vector Operations Inside the DRAM},'' in \emph{MEMSYS}, 2015.

\bibitem{lockerman2020livia}
E.~Lockerman, A.~Feldmann, M.~Bakhshalipour, A.~Stanescu, S.~Gupta, D.~Sanchez, and N.~Beckmann, ``{Livia: Data-Centric Computing Throughout the Memory Hierarchy},'' in \emph{ASPLOS}, 2020.

\bibitem{ahn2015scalable}
J.~Ahn, S.~Hong, S.~Yoo, O.~Mutlu, and K.~Choi, ``{A Scalable Processing-in-Memory Accelerator for Parallel Graph Processing},'' in \emph{ISCA}, 2015.

\bibitem{nai2017graphpim}
L.~Nai, R.~Hadidi, J.~Sim, H.~Kim, P.~Kumar, and H.~Kim, ``{GraphPIM: Enabling Instruction-Level PIM Offloading in Graph Computing Frameworks},'' in \emph{HPCA}, 2017.

\bibitem{boroumand2018google}
A.~Boroumand, S.~Ghose, Y.~Kim, R.~Ausavarungnirun, E.~Shiu, R.~Thakur, D.~Kim, A.~Kuusela, A.~Knies, P.~Ranganathan \emph{et~al.}, ``{Google Workloads for Consumer Devices: Mitigating Data Movement Bottlenecks},'' in \emph{ASPLOS}, 2018.

\bibitem{lazypim}
A.~Boroumand, S.~Ghose, B.~Lucia, K.~Hsieh, K.~Malladi, H.~Zheng, and O.~Mutlu, ``{LazyPIM: An Efficient Cache Coherence Mechanism for Processing-in-Memory},'' \emph{CAL}, 2017.

\bibitem{top-pim}
D.~Zhang, N.~Jayasena, A.~Lyashevsky, J.~L. Greathouse, L.~Xu, and M.~Ignatowski, ``{TOP-PIM: Throughput-Oriented Programmable Processing in Memory},'' in \emph{HPDC}, 2014.

\bibitem{gao2016hrl}
M.~Gao and C.~Kozyrakis, ``{HRL: Efficient and Flexible Reconfigurable Logic for Near-Data Processing},'' in \emph{HPCA}, 2016.

\bibitem{kim2018grim}
J.~S. Kim, D.~S. Cali, H.~Xin, D.~Lee, S.~Ghose, M.~Alser, H.~Hassan, O.~Ergin, C.~Alkan, and O.~Mutlu, ``{GRIM-Filter: Fast Seed Location Filtering in DNA Read Mapping Using Processing-in-Memory Technologies},'' \emph{BMC Genomics}, 2018.

\bibitem{drumond2017mondrian}
M.~Drumond, A.~Daglis, N.~Mirzadeh, D.~Ustiugov, J.~Picorel, B.~Falsafi, B.~Grot, and D.~Pnevmatikatos, ``{The Mondrian Data Engine},'' in \emph{ISCA}, 2017.

\bibitem{RVU}
P.~C. Santos, G.~F. Oliveira, D.~G. Tomé, M.~A.~Z. Alves, E.~C. Almeida, and L.~Carro, ``{Operand Size Reconfiguration for Big Data Processing in Memory},'' in \emph{DATE}, 2017.

\bibitem{NIM}
G.~F. Oliveira, P.~C. Santos, M.~A. Alves, and L.~Carro, ``{NIM: An HMC-Based Machine for Neuron Computation},'' in \emph{ARC}, 2017.

\bibitem{PEI}
J.~Ahn, S.~Yoo, O.~Mutlu, and K.~Choi, ``{PIM-Enabled Instructions: A Low-Overhead, Locality-Aware Processing-in-Memory Architecture},'' in \emph{ISCA}, 2015.

\bibitem{gao2017tetris}
M.~Gao, J.~Pu, X.~Yang, M.~Horowitz, and C.~Kozyrakis, ``{TETRIS: Scalable and Efficient Neural Network Acceleration with 3D Memory},'' in \emph{ASPLOS}, 2017.

\bibitem{Kim2016}
D.~Kim, J.~Kung, S.~Chai, S.~Yalamanchili, and S.~Mukhopadhyay, ``{Neurocube: A Programmable Digital Neuromorphic Architecture with High-Density 3D Memory},'' in \emph{ISCA}, 2016.

\bibitem{gu2016leveraging}
P.~Gu, S.~Li, D.~Stow, R.~Barnes, L.~Liu, Y.~Xie, and E.~Kursun, ``{Leveraging 3D Technologies for Hardware Security: Opportunities and Challenges},'' in \emph{GLSVLSI}, 2016.

\bibitem{boroumand2019conda}
A.~Boroumand, S.~Ghose, M.~Patel, H.~Hassan, B.~Lucia, R.~Ausavarungnirun, K.~Hsieh, N.~Hajinazar, K.~T. Malladi, H.~Zheng \emph{et~al.}, ``{CoNDA: Efficient Cache Coherence Support for Near-Data Accelerators},'' in \emph{ISCA}, 2019.

\bibitem{hsieh2016transparent}
K.~Hsieh, E.~Ebrahimi, G.~Kim, N.~Chatterjee, M.~O'Connor, N.~Vijaykumar, O.~Mutlu, and S.~W. Keckler, ``{Transparent Offloading and Mapping (TOM) Enabling Programmer-Transparent Near-Data Processing in GPU Systems},'' in \emph{ISCA}, 2016.

\bibitem{cali2020genasm}
D.~S. Cali, G.~S. Kalsi, Z.~Bing{\"o}l, C.~Firtina, L.~Subramanian, J.~S. Kim, R.~Ausavarungnirun, M.~Alser, J.~Gomez-Luna, A.~Boroumand \emph{et~al.}, ``{GenASM: A High-Performance, Low-Power Approximate String Matching Acceleration Framework for Genome Sequence Analysis},'' in \emph{MICRO}, 2020.

\bibitem{NDC_ISPASS_2014}
S.~H. Pugsley, J.~Jestes, H.~Zhang, R.~Balasubramonian \emph{et~al.}, ``{NDC: Analyzing the Impact of 3D-Stacked Memory+Logic Devices on MapReduce Workloads},'' in \emph{ISPASS}, 2014.

\bibitem{pattnaik2016scheduling}
A.~Pattnaik, X.~Tang, A.~Jog, O.~Kayiran, A.~K. Mishra, M.~T. Kandemir, O.~Mutlu, and C.~R. Das, ``{Scheduling Techniques for GPU Architectures with Processing-in-Memory Capabilities},'' in \emph{PACT}, 2016.

\bibitem{akin2015data}
B.~Akin, F.~Franchetti, and J.~C. Hoe, ``{Data Reorganization in Memory Using 3D-Stacked DRAM},'' in \emph{ISCA}, 2015.

\bibitem{hsieh2016accelerating}
K.~Hsieh, S.~Khan, N.~Vijaykumar, K.~K. Chang, A.~Boroumand, S.~Ghose, and O.~Mutlu, ``{Accelerating Pointer Chasing in 3D-Stacked Memory: Challenges, Mechanisms, Evaluation},'' in \emph{ICCD}, 2016.

\bibitem{lee2015bssync}
J.~H. Lee, J.~Sim, and H.~Kim, ``{BSSync: Processing Near Memory for Machine Learning Workloads with Bounded Staleness Consistency Models},'' in \emph{PACT}, 2015.

\bibitem{boroumand2021mitigating}
A.~Boroumand, S.~Ghose, B.~Akin, R.~Narayanaswami, G.~F. Oliveira, X.~Ma, E.~Shiu, and O.~Mutlu, ``{Mitigating Edge Machine Learning Inference Bottlenecks: An Empirical Study on Accelerating Google Edge Models},'' arXiv:2103.00768 [cs.AR], 2021.

\bibitem{boroumand2021google}
A.~Boroumand, S.~Ghose, B.~Akin, R.~Narayanaswami, G.~F. Oliveira, X.~Ma, E.~Shiu, and O.~Mutlu, ``{Google Neural Network Models for Edge Devices: Analyzing and Mitigating Machine Learning Inference Bottlenecks},'' in \emph{PACT}, 2021.

\bibitem{boroumand2022polynesia}
A.~Boroumand, S.~Ghose, G.~F. Oliveira, and O.~Mutlu, ``{Polynesia: Enabling High-Performance and Energy-Efficient Hybrid Transactional/Analytical Databases with Hardware/Software Co-Design},'' in \emph{ICDE}, 2022.

\bibitem{boroumand2021polynesia}
A.~Boroumand, S.~Ghose, G.~F. Oliveira, and O.~Mutlu, ``{Polynesia: Enabling Effective Hybrid Transactional/Analytical Databases with Specialized Hardware/Software Co-Design},'' arXiv:2103.00798 [cs.AR], 2021.

\bibitem{amiraliphd}
A.~Boroumand, ``{Practical Mechanisms for Reducing Processor-Memory Data Movement in Modern Workloads},'' Ph.D. dissertation, Carnegie Mellon University, 2020.

\bibitem{besta2021sisa}
M.~Besta, R.~Kanakagiri, G.~Kwasniewski, R.~Ausavarungnirun, J.~Ber{\'a}nek, K.~Kanellopoulos, K.~Janda, Z.~Vonarburg-Shmaria, L.~Gianinazzi, I.~Stefan \emph{et~al.}, ``{SISA: Set-Centric Instruction Set Architecture for Graph Mining on Processing-in-Memory Systems},'' in \emph{MICRO}, 2021.

\bibitem{fernandez2020natsa}
I.~Fernandez, R.~Quislant, E.~Guti{\'e}rrez, O.~Plata, C.~Giannoula, M.~Alser, J.~G{\'o}mez-Luna, and O.~Mutlu, ``{NATSA: A Near-Data Processing Accelerator for Time Series Analysis},'' in \emph{ICCD}, 2020.

\bibitem{singh2019napel}
G.~Singh, G.~, G.~F. Oliveira, S.~Corda, S.~Stuijk, O.~Mutlu, and H.~Corporaal, ``{NAPEL: Near-Memory Computing Application Performance Prediction via Ensemble Learning},'' in \emph{DAC}, 2019.

\bibitem{kwon202125}
Y.-C. Kwon, S.~H. Lee, J.~Lee, S.-H. Kwon, J.~M. Ryu, J.-P. Son, O.~Seongil, H.-S. Yu, H.~Lee, S.~Y. Kim \emph{et~al.}, ``{A 20nm 6GB Function-in-Memory DRAM, Based on HBM2 with a 1.2 TFLOPS Programmable Computing Unit using Bank-Level Parallelism, for Machine Learning Applications},'' in \emph{ISSCC}, 2021.

\bibitem{lee2021hardware}
S.~Lee, S.-h. Kang, J.~Lee, H.~Kim, E.~Lee, S.~Seo, H.~Yoon, S.~Lee, K.~Lim, H.~Shin \emph{et~al.}, ``{Hardware Architecture and Software Stack for PIM Based on Commercial DRAM Technology: Industrial Product},'' in \emph{ISCA}, 2021.

\bibitem{niu2022184qps}
D.~Niu, S.~Li, Y.~Wang, W.~Han, Z.~Zhang, Y.~Guan, T.~Guan, F.~Sun, F.~Xue, L.~Duan \emph{et~al.}, ``{184QPS/W 64Mb/$mm^2$ 3D Logic-to-DRAM Hybrid Bonding with Process-Near-Memory Engine for Recommendation System},'' in \emph{ISSCC}, 2022.

\bibitem{Sparse_MM_LiM}
Q.~Zhu, T.~Graf, H.~E. Sumbul, L.~Pileggi, and F.~Franchetti, ``{Accelerating Sparse Matrix-Matrix Multiplication with 3D-Stacked Logic-in-Memory Hardware},'' in \emph{HPEC}, 2013.

\bibitem{azarkhish2016logic}
E.~Azarkhish, C.~Pfister, D.~Rossi, I.~Loi, and L.~Benini, ``{Logic-Base Interconnect Design for Near Memory Computing in the Smart Memory Cube},'' \emph{IEEE VLSI}, 2016.

\bibitem{azarkhish2018neurostream}
E.~Azarkhish, D.~Rossi, I.~Loi, and L.~Benini, ``{Neurostream: Scalable and Energy Efficient Deep Learning with Smart Memory Cubes},'' \emph{TPDS}, 2018.

\bibitem{guo20143d}
Q.~Guo, N.~Alachiotis, B.~Akin, F.~Sadi, G.~Xu, T.~M. Low, L.~Pileggi, J.~C. Hoe, and F.~Franchetti, ``{3D-Stacked Memory-Side Acceleration: Accelerator and System Design},'' in \emph{WoNDP}, 2014.

\bibitem{de2018design}
J.~P.~C. de~Lima, P.~C. Santos, M.~A. Alves, A.~Beck, and L.~Carro, ``{Design Space Exploration for PIM Architectures in 3D-Stacked Memories},'' in \emph{CF}, 2018.

\bibitem{akin2014hamlet}
B.~Ak{\i}n, J.~C. Hoe, and F.~Franchetti, ``{HAMLeT: Hardware Accelerated Memory Layout Transform within 3D-Stacked DRAM},'' in \emph{HPEC}, 2014.

\bibitem{huang2020heterogeneous}
Y.~Huang, L.~Zheng, P.~Yao, J.~Zhao, X.~Liao, H.~Jin, and J.~Xue, ``{A Heterogeneous PIM Hardware-Software Co-Design for Energy-Efficient Graph Processing},'' in \emph{IPDPS}, 2020.

\bibitem{dai2018graphh}
G.~Dai, T.~Huang, Y.~Chi, J.~Zhao, G.~Sun, Y.~Liu, Y.~Wang, Y.~Xie, and H.~Yang, ``{GraphH: A Processing-in-Memory Architecture for Large-Scale Graph Processing},'' \emph{TCAD}, 2018.

\bibitem{liu2018processing}
J.~Liu, H.~Zhao, M.~A. Ogleari, D.~Li, and J.~Zhao, ``{Processing-in-Memory for Energy-Efficient Neural Network Training: A Heterogeneous Approach},'' in \emph{MICRO}, 2018.

\bibitem{tsai:micro:2018:ams}
P.-A. Tsai, C.~Chen, and D.~Sanchez, ``{Adaptive Scheduling for Systems with Asymmetric Memory Hierarchies},'' in \emph{MICRO}, 2018.

\bibitem{gu2020ipim}
P.~Gu, X.~Xie, Y.~Ding, G.~Chen, W.~Zhang, D.~Niu, and Y.~Xie, ``{iPIM: Programmable In-Memory Image Processing Accelerator using Near-Bank Architecture},'' in \emph{ISCA}, 2020.

\bibitem{DRAMA_CAL_2014}
A.~Farmahini-Farahani, J.~H. Ahn, K.~Compton, and N.~S. Kim, ``{DRAMA: An Architecture for Accelerated Processing Near Memory},'' \emph{Computer Architecture Letters}, 2014.

\bibitem{Asghari-Moghaddam_2016}
H.~Asghari-Moghaddam, A.~Farmahini-Farahani, K.~Morrow \emph{et~al.}, ``{Near-DRAM Acceleration with Single-ISA Heterogeneous Processing in Standard Memory Modules},'' \emph{IEEE Micro}, 2016.

\bibitem{huang2019active}
J.~Huang, R.~R. Puli, P.~Majumder, S.~Kim, R.~Boyapati, K.~H. Yum, and E.~J. Kim, ``{Active-Routing: Compute on the Way for Near-Data Processing},'' in \emph{HPCA}, 2019.

\bibitem{kersey2017lightweight}
C.~D. Kersey, H.~Kim, and S.~Yalamanchili, ``{Lightweight SIMT Core Designs for Intelligent 3D Stacked DRAM},'' in \emph{MEMSYS}, 2017.

\bibitem{li2019pims}
J.~Li, X.~Wang, A.~Tumeo, B.~Williams, J.~D. Leidel, and Y.~Chen, ``{PIMS: A Lightweight Processing-in-Memory Accelerator for Stencil Computations},'' in \emph{MEMSYS}, 2019.

\bibitem{kim2017grim}
J.~S. Kim, D.~Senol, H.~Xin, D.~Lee, S.~Ghose, M.~Alser, H.~Hassan, O.~Ergin, C.~Alkan, and O.~Mutlu, ``{GRIM-Filter: Fast Seed Filtering in Read Mapping using Emerging Memory Technologies},'' arXiv:1708.04329 [q-bio.GN], 2017.

\bibitem{boroumand2017lazypim}
A.~Boroumand, S.~Ghose, M.~Patel, H.~Hassan, B.~Lucia, N.~Hajinazar, K.~Hsieh, K.~T. Malladi, H.~Zheng, and O.~Mutlu, ``{LazyPIM: Efficient Support for Cache Coherence in Processing-in-Memory Architectures},'' arXiv:1706.03162 [cs.AR], 2017.

\bibitem{zhuo2019graphq}
Y.~Zhuo, C.~Wang, M.~Zhang, R.~Wang, D.~Niu, Y.~Wang, and X.~Qian, ``{GraphQ: Scalable PIM-Based Graph Processing},'' in \emph{MICRO}, 2019.

\bibitem{zhang2018graphp}
M.~Zhang, Y.~Zhuo, C.~Wang, M.~Gao, Y.~Wu, K.~Chen, C.~Kozyrakis, and X.~Qian, ``{GraphP: Reducing Communication for PIM-Based Graph Processing with Efficient Data Partition},'' in \emph{HPCA}, 2018.

\bibitem{lim2017triple}
H.~Lim and G.~Park, ``{Triple Engine Processor (TEP): A Heterogeneous Near-Memory Processor for Diverse Kernel Operations},'' \emph{TACO}, 2017.

\bibitem{smc_sim}
E.~Azarkhish, D.~Rossi, I.~Loi, and L.~Benini, ``{A Case for Near Memory Computation Inside the Smart Memory Cube},'' in \emph{EMS}, 2016.

\bibitem{HIVE}
M.~A.~Z. Alves, M.~Diener, P.~C. Santos, and L.~Carro, ``{Large Vector Extensions Inside the HMC},'' in \emph{DATE}, 2016.

\bibitem{jang2019charon}
J.~Jang, J.~Heo, Y.~Lee, J.~Won, S.~Kim, S.~J. Jung, H.~Jang, T.~J. Ham, and J.~W. Lee, ``{Charon: Specialized Near-Memory Processing Architecture for Clearing Dead Objects in Memory},'' in \emph{MICRO}, 2019.

\bibitem{IBM_ActiveCube}
R.~Nair, S.~F. Antao, C.~Bertolli, P.~Bose \emph{et~al.}, ``{Active Memory Cube: A Processing-in-Memory Architecture for Exascale Systems},'' \emph{IBM JRD}, 2015.

\bibitem{hadidi2017cairo}
R.~Hadidi, L.~Nai, H.~Kim, and H.~Kim, ``{CAIRO: A Compiler-Assisted Technique for Enabling Instruction-Level Offloading of Processing-in-Memory},'' \emph{TACO}, 2017.

\bibitem{santos2018processing}
P.~C. Santos, G.~F. Oliveira, J.~P. Lima, M.~A. Alves, L.~Carro, and A.~C. Beck, ``{Processing in 3D Memories to Speed Up Operations on Complex Data Structures},'' in \emph{DATE}, 2018.

\bibitem{Chi2016}
P.~Chi, S.~Li, C.~Xu, T.~Zhang, J.~Zhao, Y.~Liu, Y.~Wang, and Y.~Xie, ``{PRIME: A Novel Processing-in-Memory Architecture for Neural Network Computation in ReRAM-Based Main Memory},'' in \emph{ISCA}, 2016.

\bibitem{Shafiee2016}
A.~Shafiee, A.~Nag, N.~Muralimanohar, R.~Balasubramonian, J.~P. Strachan, M.~Hu, R.~S. Williams, and V.~Srikumar, ``{ISAAC: A Convolutional Neural Network Accelerator with In-Situ Analog Arithmetic in Crossbars},'' in \emph{ISCA}, 2016.

\bibitem{seshadri2017ambit}
V.~Seshadri, D.~Lee, T.~Mullins, H.~Hassan, A.~Boroumand, J.~Kim, M.~A. Kozuch, O.~Mutlu, P.~B. Gibbons, and T.~C. Mowry, ``{Ambit: In-Memory Accelerator for Bulk Bitwise Operations Using Commodity DRAM Technology},'' in \emph{MICRO}, 2017.

\bibitem{seshadri2019dram}
V.~Seshadri and O.~Mutlu, ``{In-DRAM Bulk Bitwise Execution Engine},'' arXiv:1905.09822 [cs.AR], 2019.

\bibitem{li2017drisa}
S.~Li, D.~Niu, K.~T. Malladi, H.~Zheng, B.~Brennan, and Y.~Xie, ``{DRISA: A DRAM-Based Reconfigurable In-Situ Accelerator},'' in \emph{MICRO}, 2017.

\bibitem{seshadri2013rowclone}
V.~Seshadri, Y.~Kim, C.~Fallin, D.~Lee, R.~Ausavarungnirun, G.~Pekhimenko, Y.~Luo, O.~Mutlu, P.~B. Gibbons, M.~A. Kozuch \emph{et~al.}, ``{RowClone: Fast and Energy-Efficient In-DRAM Bulk Data Copy and Initialization},'' in \emph{MICRO}, 2013.

\bibitem{seshadri2016processing}
V.~Seshadri and O.~Mutlu, ``{The Processing Using Memory Paradigm: In-DRAM Bulk Copy, Initialization, Bitwise AND and OR},'' arXiv:1610.09603 [cs.AR], 2016.

\bibitem{deng2018dracc}
Q.~Deng, L.~Jiang, Y.~Zhang, M.~Zhang, and J.~Yang, ``{DrAcc: A DRAM Based Accelerator for Accurate CNN Inference},'' in \emph{DAC}, 2018.

\bibitem{xin2020elp2im}
X.~Xin, Y.~Zhang, and J.~Yang, ``{ELP2IM: Efficient and Low Power Bitwise Operation Processing in DRAM},'' in \emph{HPCA}, 2020.

\bibitem{song2018graphr}
L.~Song, Y.~Zhuo, X.~Qian, H.~Li, and Y.~Chen, ``{GraphR: Accelerating Graph Processing Using ReRAM},'' in \emph{HPCA}, 2018.

\bibitem{song2017pipelayer}
L.~Song, X.~Qian, H.~Li, and Y.~Chen, ``{PipeLayer: A Pipelined ReRAM-Based Accelerator for Deep Learning},'' in \emph{HPCA}, 2017.

\bibitem{gao2019computedram}
F.~Gao, G.~Tziantzioulis, and D.~Wentzlaff, ``{ComputeDRAM: In-Memory Compute Using Off-the-Shelf DRAMs},'' in \emph{MICRO}, 2019.

\bibitem{eckert2018neural}
C.~Eckert, X.~Wang, J.~Wang, A.~Subramaniyan, R.~Iyer, D.~Sylvester, D.~Blaauw, and R.~Das, ``{Neural Cache: Bit-Serial In-Cache Acceleration of Deep Neural Networks},'' in \emph{ISCA}, 2018.

\bibitem{aga2017compute}
S.~Aga, S.~Jeloka, A.~Subramaniyan, S.~Narayanasamy, D.~Blaauw, and R.~Das, ``{Compute Caches},'' in \emph{HPCA}, 2017.

\bibitem{dualitycache}
D.~Fujiki, S.~Mahlke, and R.~Das, ``{Duality Cache for Data Parallel Acceleration},'' in \emph{ISCA}, 2019.

\bibitem{seshadri2016buddy}
V.~Seshadri, D.~Lee, T.~Mullins, H.~Hassan, A.~Boroumand, J.~Kim, M.~A. Kozuch, O.~Mutlu, P.~B. Gibbons, and T.~C. Mowry, ``{Buddy-RAM: Improving the Performance and Efficiency of Bulk Bitwise Operations Using DRAM},'' arXiv:1611.09988 [cs.AR], 2016.

\bibitem{seshadri.bookchapter17}
V.~Seshadri and O.~Mutlu, ``{Simple Operations in Memory to Reduce Data Movement},'' in \emph{Advances in Computers, Volume 106}, 2017.

\bibitem{seshadri2018rowclone}
V.~Seshadri, Y.~Kim, C.~Fallin, D.~Lee, R.~Ausavarungnirun, G.~Pekhimenko, Y.~Luo, O.~Mutlu, P.~B. Gibbons, M.~A. Kozuch \emph{et~al.}, ``{RowClone: Accelerating Data Movement and Initialization Using DRAM},'' arXiv:1805.03502 [cs.AR], 2018.

\bibitem{seshadri2015fast}
V.~Seshadri, K.~Hsieh, A.~Boroum, D.~Lee, M.~A. Kozuch, O.~Mutlu, P.~B. Gibbons, and T.~C. Mowry, ``{Fast Bulk Bitwise AND and OR in DRAM},'' \emph{CAL}, 2015.

\bibitem{li2016pinatubo}
S.~Li, C.~Xu, Q.~Zou, J.~Zhao, Y.~Lu, and Y.~Xie, ``{Pinatubo: A Processing-in-Memory Architecture for Bulk Bitwise Operations in Emerging Non-Volatile Memories},'' in \emph{DAC}, 2016.

\bibitem{ferreira2021pluto}
J.~D. Ferreira, G.~Falcao, J.~G{\'o}mez-Luna, M.~Alser, L.~Orosa, M.~Sadrosadati, J.~S. Kim, G.~F. Oliveira, T.~Shahroodi, A.~Nori \emph{et~al.}, ``{pLUTo: In-DRAM Lookup Tables to Enable Massively Parallel General-Purpose Computation},'' arXiv:2104.07699 [cs.AR], 2021.

\bibitem{ferreira2022pluto}
J.~D. Ferreira, G.~Falcao, J.~G{\'o}mez-Luna, M.~Alser, L.~Orosa, M.~Sadrosadati, J.~S. Kim, G.~F. Oliveira, T.~Shahroodi, A.~Nori \emph{et~al.}, ``{pLUTo: Enabling Massively Parallel Computation in DRAM via Lookup Tables},'' in \emph{MICRO}, 2022.

\bibitem{imani2019floatpim}
M.~Imani, S.~Gupta, Y.~Kim, and T.~Rosing, ``{FloatPIM: In-Memory Acceleration of Deep Neural Network Training with High Precision},'' in \emph{ISCA}, 2019.

\bibitem{he2020sparse}
Z.~He, L.~Yang, S.~Angizi, A.~S. Rakin, and D.~Fan, ``{Sparse BD-Net: A Multiplication-Less DNN with Sparse Binarized Depth-Wise Separable Convolution},'' \emph{JETC}, 2020.

\bibitem{flashcosmos}
J.~Park, R.~Azizi, G.~F. Oliveira, M.~Sadrosadati, R.~Nadig, D.~Novo, J.~G{\'o}mez-Luna, M.~Kim, and O.~Mutlu, ``{Flash-Cosmos: In-Flash Bulk Bitwise Operations Using Inherent Computation Capability of NAND Flash Memory},'' in \emph{MICRO}, 2022.

\bibitem{truong2022adapting}
M.~S. Truong, L.~Shen, A.~Glass, A.~Hoffmann, L.~R. Carley, J.~A. Bain, and S.~Ghose, ``{Adapting the RACER Architecture to Integrate Improved In-ReRAM Logic Primitives},'' \emph{JETCAS}, 2022.

\bibitem{truong2021racer}
M.~S. Truong, E.~Chen, D.~Su, L.~Shen, A.~Glass, L.~R. Carley, J.~A. Bain, and S.~Ghose, ``{RACER: Bit-Pipelined Processing Using Resistive Memory},'' in \emph{MICRO}, 2021.

\bibitem{olgun2021quactrng}
A.~Olgun, M.~Patel, A.~G. Ya\u{g}l{\i}k\c{c}{\i}, H.~Luo, J.~S. Kim, F.~N. Bostanc{\i}, N.~Vijaykumar, O.~Ergin, and O.~Mutlu, ``{QUAC-TRNG: High-Throughput True Random Number Generation Using Quadruple Row Activation in Commodity DRAMs},'' in \emph{ISCA}, 2021.

\bibitem{kim2019d}
J.~S. Kim, M.~Patel, H.~Hassan, L.~Orosa, and O.~Mutlu, ``{D-RaNGe: Using Commodity DRAM Devices to Generate True Random Numbers With Low Latency and High Throughput},'' in \emph{HPCA}, 2019.

\bibitem{kim2018dram}
J.~S. Kim, M.~Patel, H.~Hassan, and O.~Mutlu, ``{The DRAM Latency PUF: Quickly Evaluating Physical Unclonable Functions by Exploiting the Latency-Reliability Tradeoff in Modern Commodity DRAM Devices},'' in \emph{HPCA}, 2018.

\bibitem{bostanci2022dr}
F.~N. Bostanc{\i}, A.~Olgun, L.~Orosa, A.~G. Ya{\u{g}}l{\i}k{\c{c}}{\i}, J.~S. Kim, H.~Hassan, O.~Ergin, and O.~Mutlu, ``{DR-STRaNGe: End-to-End System Design for DRAM-Based True Random Number Generators},'' in \emph{HPCA}, 2022.

\bibitem{olgun2022pidram}
A.~Olgun, J.~G. Luna, K.~Kanellopoulos, B.~Salami, H.~Hassan, O.~Ergin, and O.~Mutlu, ``{PiDRAM: A Holistic End-to-End FPGA-Based Framework for Processing-in-DRAM},'' \emph{TACO}, 2022.

\bibitem{ali2019memory}
M.~F. Ali, A.~Jaiswal, and K.~Roy, ``{In-Memory Low-Cost Bit-Serial Addition Using Commodity DRAM Technology},'' in \emph{{TCAS-I}}, 2019.

\bibitem{angizi2019graphide}
S.~Angizi and D.~Fan, ``{GraphiDe: A Graph Processing Accelerator Leveraging In-DRAM-Computing},'' in \emph{GLSVLSI}, 2019.

\bibitem{li2018scope}
S.~Li, A.~O. Glova, X.~Hu, P.~Gu, D.~Niu, K.~T. Malladi, H.~Zheng, B.~Brennan, and Y.~Xie, ``{SCOPE: A Stochastic Computing Engine for DRAM-Based In-Situ Accelerator},'' in \emph{MICRO}, 2018.

\bibitem{subramaniyan2017parallel}
A.~Subramaniyan and R.~Das, ``{Parallel Automata Processor},'' in \emph{ISCA}, 2017.

\bibitem{zha2020hyper}
Y.~Zha and J.~Li, ``{Hyper-AP: Enhancing Associative Processing Through A Full-Stack Optimization},'' in \emph{ISCA}, 2020.

\bibitem{fujiki2018memory}
D.~Fujiki, S.~Mahlke, and R.~Das, ``{In-Memory Data Parallel Processor},'' in \emph{ASPLOS}, 2018.

\bibitem{orosa2021codic}
L.~Orosa, Y.~Wang, M.~Sadrosadati, J.~Kim, M.~Patel, I.~Puddu, H.~Luo, K.~Razavi, J.~G{\'o}mez-Luna, H.~Hassan, N.~M. Ghiasi, S.~Ghose, and O.~Mutlu, ``{CODIC: A Low-Cost Substrate for Enabling Custom In-DRAM Functionalities and Optimizations},'' in \emph{ISCA}, 2021.

\bibitem{sharad2013ultra}
M.~Sharad, D.~Fan, and K.~Roy, ``{Ultra Low Power Associative Computing with Spin Neurons and Resistive Crossbar Memory},'' in \emph{DAC}, 2013.

\bibitem{rezaei2020nom}
S.~H.~S. Rezaei, M.~Modarressi, R.~Ausavarungnirun, M.~Sadrosadati, O.~Mutlu, and M.~Daneshtalab, ``{NoM: Network-on-Memory for Inter-Bank Data Transfer in Highly-Banked Memories},'' \emph{CAL}, 2020.

\bibitem{hajinazarsimdram}
N.~Hajinazar, G.~F. Oliveira, S.~Gregorio, J.~D. Ferreira, N.~M. Ghiasi, M.~Patel, M.~Alser, S.~Ghose, J.~G{\'o}mez-Luna, and O.~Mutlu, ``{SIMDRAM: A Framework for Bit-Serial SIMD Processing Using DRAM},'' in \emph{ASPLOS}, 2021.

\bibitem{chang2016low}
K.~K. Chang, P.~J. Nair, D.~Lee, S.~Ghose, M.~K. Qureshi, and O.~Mutlu, ``{Low-Cost Inter-Linked Subarrays (LISA): Enabling Fast Inter-Subarray Data Movement in DRAM},'' in \emph{HPCA}, 2016.

\bibitem{devicethree}
T.~K.~D. Community", ``{Linux and the Devicetree — The Linux Kernel Documentation},'' \url{https://www.kernel.org/doc/html/latest/devicetree/usage-model.html}.

\bibitem{kim2020revisiting}
J.~S. Kim, M.~Patel, A.~G. Ya{\u{g}}l{\i}k{\c{c}}{\i}, H.~Hassan, R.~Azizi, L.~Orosa, and O.~Mutlu, ``{Revisiting RowHammer: An Experimental Analysis of Modern DRAM Devices and Mitigation Techniques},'' in \emph{ISCA}, 2020.

\bibitem{orosa2021deeper}
L.~Orosa, A.~G. Yaglikci, H.~Luo, A.~Olgun, J.~Park, H.~Hassan, M.~Patel, J.~S. Kim, and O.~Mutlu, ``{A Deeper Look into RowHammer's Sensitivities: Experimental Analysis of Real DRAM Chips and Implications on Future Attacks and Defenses"},'' in \emph{MICRO}, 2021.

\bibitem{yauglikcci2022understanding}
A.~G. Ya{\u{g}}l{\i}k{\c{c}}{\i}, H.~Luo, G.~F. De~Oliviera, A.~Olgun, M.~Patel, J.~Park, H.~Hassan, J.~S. Kim, L.~Orosa, and O.~Mutlu, ``{Understanding RowHammer Under Reduced Wordline Voltage: An Experimental Study Using Real DRAM Devices},'' in \emph{DSN}, 2022.

\bibitem{knowlton1966programmer}
K.~C. Knowlton, ``{A Programmer's Description of L6},'' \emph{CACM}, 1966.

\bibitem{johnson1973near}
D.~S. Johnson, ``{Near-Optimal Bin Packing Algorithms},'' Ph.D. dissertation, Massachusetts Institute of Technology, 1973.

\bibitem{qemu}
F.~Bellard, ``{QEMU, a Fast and Portable Dynamic Translator},'' in \emph{USENIX ATC}, 2005.

\end{thebibliography}


\end{document}